\documentclass[final,3p,times]{elsarticle}
\usepackage{algorithm,algorithmicx,algpseudocode}

\pdfoutput=1
\usepackage{graphicx}
\usepackage{epsfig}
\usepackage{amsmath}

\journal{Computer Physics Communications}

\begin{document}

\begin{frontmatter}

\title{Large-scale Ferrofluid Simulations on Graphics Processing Units}

\author[SSU,UniAug]{A.~Yu.~Polyakov}
\author[SSU]{T.~V.~Lyutyy}
\author[UniAug]{S.~Denisov}
\author[SSU]{V.~V.~Reva}
\author[UniAug]{P.~H\"anggi}

\address[SSU]{Sumy State University, Sumy, Ukraine}
\address[UniAug]{Institute of Physics, University of Augsburg, Augsburg,
Germany}


\begin{abstract}
We present an approach to
molecular-dynamics simulations of ferrofluids on graphics processing units (GPUs).
Our numerical scheme
is based on a GPU-oriented modification of the Barnes-Hut (BH) algorithm designed
to increase the parallelism of computations.
For an ensemble consisting of one million of ferromagnetic
particles, the performance of the proposed algorithm on a Tesla M2050 GPU
demonstrated a computational-time speed-up of four order of magnitude compared to the performance of the
sequential All-Pairs (AP) algorithm on a single-core CPU, and two order of magnitude compared to the performance of the
optimized AP algorithm on the GPU. The accuracy of the
scheme is corroborated by comparing the results of numerical simulations with theoretical predictions.
\end{abstract}

\begin{keyword}
Ferrofluid, molecular dynamics simulation, the Barnes-Hut algorithm, GPUs, CUDA
\end{keyword}
\end{frontmatter}

\section{Introduction}
Ferrofluids are media composed of magnetic nanoparticles of diameters in the range $10 \div 50$ nm which are
dispersed in a viscous fluid (for example, water or ethylene glycol) \cite{Rosensweig}. These
physical systems combine the basic properties of liquids, i.e. a viscosity and the presence of surface
tension, and those of ferromagnetic solids possessing, i.e. an
internal magnetization and high permeability to magnetic fields. This synergetic duality makes
ferrofluids an attractive candidate for performing different tasks, ranging from the delivery of rocket fuel into spacecraft's
thrust chambers under zero-gravity conditions to the high-precision drug delivery during cancer therapy \cite{Pankhrust2003}.
Moreover, ferrofluids have already found their way into commercial applications as EM-controlled shock absorbers, 
dynamic seals in engines and computer hard-drives,
and as key elements of high-quality loudspeakers, to name but a few \cite{Raj1990}.

Being liquids, ferrofluids can be modeled by using different
macroscopic, continuum-media approaches. The corresponding field, ferrohydrodynamics \cite{Rosensweig}, is a well-established area that has
produced a lot of fundamental results. However, some important phenomena such as magnetoviscosity
\cite{McTague1969, Odenbach2002_book} cannot be described properly neither on the hydrodynamical level nor within the single-particle
picture \cite{Odenbach2000}. The role of multiparticle aggregates -- chains and clusters -- living in the ferrofluid bulk,
is important there. The evaluation of other non-Newtonian features of ferrofluids, which can be controlled by external
magnetic fields \cite{Odenbach2002_book}, also demands information about structure of aggregates and their dynamics.
The key mechanism responsible for the formation of ferro-clusters is the dipole-dipole interaction acting between magnetic particles. 
Under certain conditions, the interaction effects can overcome the destructive effects of thermal fluctuations and contribute tangibly 
to the ensemble dynamics. Therefore, in order to get a deeper insight
into the above-mentioned features of ferrofluids, the interaction effects should be explicitly included into the model.

Strictly speaking, the dipole-dipole interaction acts between all the possible pairs of $N$ ferromagnetic particles.
Therefore, the computational time of the straightforward sequential algorithm scales like $N^2$. This is a fundamental drawback of
many-body simulations and standard, CPU-based computational resources often limit the scales of the molecular-dynamics calculations.
In order to run larger ensembles for longer times, researchers either (i) advance their models and numerical schemes and/or (ii) rely on more efficient computers. The first track led to different methods developed to explore
the equilibrium properties of bulk ferrofluids. Most prominent are
cut-off sphere approximations \cite{Ilg2006} and the Ewald summation technique
\cite{deLeeuw1980}, see also Ref. \cite{Cerda2008}
for different modifications of the technique. However, both methods are of limited use for computational studies of
confined ferrofluids, a problem that attracts special attention nowadays due to its practical relevance \cite{Hartshorne2004, Pamme2006}. 
An alternative approach
is the search for ways to increase scalability and parallelism of computations, see Ref. \cite{phil2005}.
Until recently, this practically meant the use of either large computational clusters, consisting of many Central Processing Units
(CPUs), or massively parallel supercomputers, like Blue Gene supercomputers. The prices and the maintenance costs
of such devices are high. The advent of the general-purpose computing on graphics processing units (GP$^2$U) \cite{Harris2005}
has changed the situation drastically
and boosted simulations of many-body systems onto new level \cite{Nyland2007}. GPUs, initially designed to serve as the data pipelines
for graphical information, are relatively small, much easier to maintain, and possess high computation capabilities allowing for parallel data 
processing. Nowadays, the scientific GPU-computing is used in many areas of computational
physics, thanks to the Compute Unified Device Architecture (CUDA) developed by NVIDIA Corporation \cite{Sanders2011}. CUDA significantly
simplifies GPU-based calculations so now one could use a Sony PlayStation 3 as a multi-core computer by programming it with $C$, $C++$ or
$Fortran$ \cite{CUDAfort} languages.

The typical scale of molecular-dynamics simulations in ferrofluid studies is within
the range $N = 10^2 - 10^3$ \cite{Wang2002, AA2007}, while $N = 10^4$ constitutes the current limit \cite{Cerda2010}.
It is evident that the increase of the ensemble size by several orders of magnitude would tangibly
improve the statistical sampling and thus the quality of simulation results. To run numerically $10^5 \div 10^6$ interacting
magnetic particles is a task not much different from the running of an
artificial Universe \cite{Springel2005}. Therefore, similar to the case of Computational Cosmology \cite{Zwart2007, Belleman2008, Aubert2010},
the GP$^2$U seems to be very promising also in the context of ferrofluid simulations. In this paper we report the performance of
a recently proposed GPU-oriented modification of
the Barnes-Hut algorithm \cite{Burtscher2011} used to simulate the dynamics of $N = 10^3 \div 10^6$ interacting ferromagnetic
particles moving in a viscous medium.

Although being mentioned as a potentially promising approach (see, for example, Ref. \cite{Holm2004}),
the Barnes-Hut algorithm was never used in molecular-dynamics ferrofluid simulations, to the best of our knowledge.
Our main goal here is to demonstrate a sizable speed-up one can reach when modeling such systems on a GPU -- \textit{and} by using 
GPU-oriented numerical algorithms -- compared to the performance of conventional, CPU-based algorithms. We also demonstrate the high accuracy 
of the Barnes-Hut approximation by using several benchmarks. The paper is organized as follows: First, in Section
II we specify the model. Then, in Section III, we describe how both, the All-Pairs and
Barn-Hut algorithms, can be efficiently implemented for GPU-based simulations. The results of numerical tests are presented and discussed 
in Section V. Finally, Section VI contains conclusions.

\section{The model}
The model system represents an ensemble of $N$ identical
particles of the radius
$R$, made of a ferromagnetic material of density $D$ and 
specific magnetization $\mu$. Each particle occupies volume $V
= \frac{3}{4}\pi R ^3$, has magnetic moment $\vec{m} = \vec{m}(t)$ of
constant magnitude $|\vec{m}| = m = V \mu $, mass $M = V D$, and moment
of inertia $I = \frac{2}{5} M R^2$. The ensemble is dispersed in a liquid of
viscosity $\eta$. Based on the Langevin dynamics approach, the equations of
motions for $k$-th nanoparticle can be written in the following form
\cite{Wang2002}
\begin{eqnarray}
I\ddot{\theta_k} \!&=&\! N_{kz} - G_r \dot \theta_k + \Xi^{r}_{\theta },
\\[6 pt]
\label{eq:mot_rot_tet}
I\ddot{\varphi_k} \!&=&\! - N_{kx} \sin \varphi_k + N_{ky} \cos \varphi_k -
G_r \dot \varphi_k + \Xi^{r}_{\varphi},
\\[6 pt]
\label{eq:mot_rot_phi}
M\ddot {\vec r}_k \!&=&\! \mu_0(\vec m_k \nabla_k) \cdot \vec H_k
+ \vec F_k^{sr} - G_d \dot {\vec r}_k
+ \vec{\Xi}^{d}_{\vec r},
\label{eq:red_mot_disp}
\end{eqnarray}
where $\theta$ and $\varphi$ are the polar and azimuthal angles of the
magnetization vector $\vec{m}$ respectively. $N_{kx} = m_{ky} H_{kz} - m_{kz} H_{ky}, N_{ky} =
m_{kz} H_{kx} - m_{kx} H_{kz}, N_{kz} = m_{ky} H_{kx} - m_{kx} H_{ky}$, $x, y,
z$ denotes the Cartesian coordinates, dots over the variables denote the
derivatives with respect to time. $\vec{r}_k$ is the
radius-vector defining the nanoparticle position, and the gradient is given by $\nabla_k = \frac
{\partial} {\partial \vec{r}_k} = \vec e_x \frac {\partial}{\partial x_k} + \vec
e_y \frac {\partial}{\partial y_k} + \vec e_z \frac {\partial}{\partial z_k}$,
($\vec e_x, \vec e_y, \vec e_z$ are the unit vectors of the Cartesian
coordinates). Constants $G_t = 6 \pi \eta R$, and $G_r = 8 \pi \eta R
^3$ specify translational and rotational friction coefficients, $\mu_0 = 4 \pi \cdot
10^{-7} \mathrm{H}/\mathrm{m}$ is the magnetic constant.

The resulting field acting on the $k$-th particle 
$\vec H_k$ is the sum of external
field
$\vec H^{ext}$ and overall field exerted on the particle by the rest of the
ensemble,
\begin{eqnarray}
\vec H_k \!&=&\! \sum_{j = 1, j \neq k}^{N}{\vec H_{kj}^{dip}} + \vec
H^{ext},
\\[6 pt]
\label{eq:H}
\vec H_{kj}^{dip} \!&=&\! \frac{3 \vec r_{kj} (\vec m_j \vec r_{kj})
- \vec m_j {\vec r_{kj}} ^{\,2}} {|\vec r_{kj}| ^5},
\label{eq:H_dip}
\end{eqnarray}
where $\vec r_{kj} = \vec r_k - \vec r_j$. $\vec F_k^{sr}$ in
Eq.~\ref{eq:red_mot_disp} denotes the force
induced by a short-range interaction
potential. In this paper we use Lennard-Jones potential \cite{Wang2002}
(though hard sphere
\cite{Weis1993}, soft sphere \cite{Wei1992} and Yukawa-type \cite{Meriguet2005}
potentials can be used as alternatives), so that
\begin{equation}
\vec F_k^{sr} = 24E \sum_{j = 1, j \neq k}^{N}
{\frac{\vec r_{kj}} {\vec r_{kj} ^{\,2}}
\left[{{\left(\frac{s}{\vec r_{kj}}\right)} ^{12} -
\left(\frac {s} {\vec r_{kj}} \right) ^6} \right]}.
\label{eq:LJ}
\end{equation}
Here $E$ is the depth of the potential well and $s$ is the equilibrium distance at
which the inter-particle force vanishes. The interaction between a particle
and container walls is also modeled with a Lennard-Jones potential of the same type.
The random-force vector,
representing the interaction of a particle with thermal bath, has standard white-noise components,
$\langle \Xi_\alpha^r (t) \rangle = 0$ ($\alpha = \varphi, \theta$),
$\langle \Xi_\beta^d (t) \rangle = 0$ ($\beta = x, y, z)$,
and second moments satisfying
$\langle \Xi_{r\alpha} (t) \Xi_{r\beta} (t') \rangle
=3 k_{\rm B} T G_r \delta_{\alpha \beta} \delta (t - t')\
(\alpha, \beta = \varphi, \theta )$
$\langle \Xi_{d \alpha} (t) \Xi_{d \beta} (t') \rangle =
2 k_{\rm B} T G_t \delta_{\alpha \beta } \delta
(t - t') \ (\alpha ,\beta = x, y, z)$ \cite{Wang2002}.
Here $k_{\rm B}$ is
the Boltzmann constant and $T$ is the temperature of the heat bath.

By rescaling the variables, $\tau = t / T_{ch},
(T_{ch} = R / \mu \sqrt {3 D / 4 \mu_0}),
\vec u = \vec m / |\vec m|, \vec \rho_k = \vec r_k / R$,
${\nabla_{\rho_k}} = \frac{1}{R} \frac{d}{d \vec \rho_k}$,
the equations of motions for the $k-$th particle can be rewritten in the reduced form, 
\begin{eqnarray}
\frac{2}{5} \cdot \frac{d^2 \theta_k}{d \tau^2} \!&=&\!
(u_{kx} h_{ky} - u_{ky} h_{kx}) - \varGamma_r \frac{d \theta_k}{d \tau} +
\frac{T_{ch} ^2}{V R ^2 D} \xi^r_\theta \
\label{eq:red_mot_rot_tet}
\\
\frac{2}{5} \cdot \frac{d ^2 \varphi_k}{d \tau ^2} \!&=&\! -(u_{ky} h_{kz} -
u_{kz} h_{ky})
\sin \varphi_k + (u_{kz} h_{kx} - u_{kx} h_{kz}) \cos \varphi_k
- \varGamma_r \frac{d \varphi_k}{d \tau} +
\frac{T_{ch} ^2}{V R ^2 D}{\xi^r_\varphi}\
\label{eq:red_mot_rot_phi}
\\
\frac{d^2 \vec{\rho}_k}{d \tau^2} \!&=&\! \vec f_k^{dip} + \vec f_k^{sr}
- \varGamma_d \frac{d \vec \rho_k}{d \tau}
+ \frac{T_{ch}^2}{V R D}{\vec \xi_{\vec \rho}^d}
\label{eq:red_mot_disp_final}
\end{eqnarray}
where
\begin{eqnarray}
\vec h_k \!&=&\! \sum_{j = 1, j \neq k}^{N}{\vec h_{kj}^{dip}} + \vec
h^{ext},
\label{eq:h_red}
\\
\vec h_{kj}^{dip} \!&=&\! \frac { 3 \vec \rho_{kj} (\vec u_j
\vec \rho_{kj}) - \vec u_j \vec \rho_{kj} ^{\, 2}} {\rho_{kj} ^{\, 5}},
\label{eq:h_dip_red}
\end{eqnarray}

\begin{eqnarray}
\vec f_k^{dip} \!&=&\! \sum_{j = 1, j \neq k}^{N} \left[
3 \frac {\vec \rho_{kj} (\vec u_j \vec u_k) + \vec u_k (\vec u_j \vec
\rho_{kj}) + \vec u_j (\vec u_k \vec \rho_{kj})} {\rho_{kj}^{\, 5}} -
15 \frac {\vec \rho_{kj} (\vec u_k \vec \rho_{kj})(\vec u_j \vec \rho_{kj})}
{\rho_{kj}^{\, 7}}
\right ],
\end{eqnarray}

\begin{eqnarray}
\vec f_k^{sr} \!&=&\! 24 \varepsilon \sum_{j = 1, j \neq k}^{N}
\frac {\vec \rho_{kj}} {\vec \rho_{kj} ^{\, 2}}
\left[ {\left( \frac{\sigma}{\rho_{kj}} \right)} ^{12}
- {\left (\frac{\sigma}{\rho_{kj}} \right) } ^6 \right],
\end{eqnarray}\
Here
$\vec h^{ext} = 3 \vec H^{ext} / 4 \pi \mu,
\varGamma_t = G_t T_{ch} / V D,
\varGamma_r = G_r T_{ch} / V R^2 D,
\sigma = s / R, \varepsilon = E {T_{ch}}^2 / V R D,
\vec \rho_{kj} = \vec \rho_k - \vec \rho_j$

Random-force vector components are given now by white Gaussian noises, with the second moments
satisfying
$\langle {\xi_\alpha^r}(\tau){\xi_\beta^r}(\tau')\rangle
= 3 k_{\rm B} T \varGamma_r \delta_{\alpha \beta} \delta (\tau - \tau') /
T_{ch}$
$(\alpha, \beta = \varphi, \theta)$
$\langle \xi_\alpha^d (\tau) \xi_\beta^d (\tau') \rangle
= 2 k_{\rm B} T \varGamma_t \delta_{\alpha \beta } \delta (\tau -\tau') /
T_{ch}$
$(\alpha, \beta = x, y, z)$. In general case
the characteristic relaxation time of particle magnetic moments to their equilibrium orientations
is much smaller than $T_{ch}$, and one can neglect
the magnetization dynamics by assuming that the
direction of the vector $\vec{m}_k$ coincides with the easy axis of the $k-$th particle.
We assume that this condition holds for our model.

\section{Two approaches to many-body simulations on GPUs: the All-Pairs and the Barnes-Hut algorithms}

In this section we discuss two alternative approaches to the numerical propagation of the dynamical system given by Eqs.
(\ref{eq:red_mot_rot_tet} - \ref{eq:red_mot_disp_final}) on a GPU. It is assumed that the reader is familiar with the basics of
the GP$^2$U, otherwise we address him to Refs. (\cite{Januszewski2010, Weigel2011}) that contain crash-course-like introductions 
into the physically-oriented GPU computing.

\subsection{All-Pair Algorithm}

The most straightforward approach to propagate a system of $N$ interacting particles is to account for the interactions between all pairs.
Although exact, the corresponding All-Pairs (AP) algorithm is slow when performed on a CPU, so it is usually used to propagate systems
of $N = 10^2 \div 10^3$ particles. However, even with this brute-force method one can tangibly benefit
from GPU computations by noticing that the AP idea fits CUDA architecture \cite{GpuGems3}.

One integration step of the standard AP algorithm is performed in two stages. They are:
\begin{enumerate}
 \item {Calculation of increment of particle positions and magnetic moments
directions.}
 \item {Update particle positions and magnetic moments directions. }
\end{enumerate}

This structure remains intact in the GPU version of the algorithm. 
Kernels responsible for the first stage compute forces that act on
the particles, and calculate the corresponding increments for particle
positions and magnetic moments, according to equations
(\ref{eq:red_mot_rot_tet} - \ref{eq:red_mot_disp_final}). The increments are then written
into the global memory. Finally, second-stage kernels update particle states with the
obtained increments. There is, however, a need for the
global synchronization of the threads that belong to the different blocks after every stage 
since the stages are performed on separate CUDA kernels and all the information about the state of the 
system is kept in the shared memory. 

Each thread is responsible for one particle of the ensemble, and thus
it should account for the forces exerted on the particle by the rest of the ensemble.
To speed up the computational process we keep the data vector of the thread
particle in the shared memory, as well as the information
on other particles, needed to compute the corresponding interaction forces. Thus we have
two sets of arrays of data in the shared memory, namely

\begin{enumerate}
 \item {data of the particles assigned to the threads of the block;}
 \item {data of particles to compute interaction with.}
\end{enumerate}

The necessary data are the coordinates of the particles and
projections of their magnetic moment vectors onto $x$, $y$ and $z$ axis. Size of
the arrays are equal to the number of threads per block. The first set
of the data is constant during one integration step, but the second set is
changed. So, at the beginning we upload the information on particle coordinates and momenta
to the second set of arrays in the shared memory. After computing the forces acting
on the block particles, the procedure is repeated, i. e.
the information on another set of particles is written into the second set of arrays
and the corresponding interactions are computed. The corresponding pseudo-code
is presented the below\footnote {In the pseudo-code $\varDelta \rho_x$, $\varDelta \rho_y$, $\varDelta \rho_z$,
$\varDelta \theta$, $\varDelta \varphi$ denote increments of a particle's
coordinates $x$, $y$, $z$, and the direction angles of particle magnetic moment, $\theta$ and
$\varphi$, needed to propagate the particle over one time step $\varDelta \tau$.}.

\begin{algorithm*}
\caption{A CUDA kernel that computes increments.}
\label{alg:AllBodies}
\begin{algorithmic}[1]
\State cached $k \leftarrow blockIdx.x \cdot blockDim.x + threadIdx.x$
\For{$j = 0$ to $numberOfParticles$}
\State upload to the shared memory particle positions and angles
\State e.g. $x_{shared}[threadIdx.x] = x_{global}[ind]$, etc.
\For{$i = 0$ to $blockDim.x$}
\If{$j + i \neq k$ }	
\Comment{the condition to avoid particle on-self influence}
\State Calculate force and dipole field for the particle number $j + i$
\State on current $k$-th particle, and add the result to
\State the total cached $\vec f^{dip}$, $\vec f^{sr}$, $\vec h^{dip}$.
\EndIf
\EndFor
\State \_\_syncthreads();
\State $j \leftarrow j + blockDim.x$
\EndFor
\State Update cached $\varDelta \rho_x$, $\varDelta \rho_y$, $\varDelta
\rho_z$, $\varDelta \theta$, $\varDelta \varphi$ according to equations
\ref{eq:red_mot_rot_tet} -
\ref{eq:red_mot_disp_final}.
\State Copy increments to global memory.
\end{algorithmic}
\end{algorithm*}

The advantage of the described approach is that it uses the global memory in the most optimal way.
The access to global memory is coalesced and there are no shared
memory bank conflicts. For an ensemble $N = 10^4$ particles this leads to the GPU occupancy 
\footnote{The parameter shows how the GPU is kept busy, and it is equal to the ratio of the number of active
warps to the maximum number of warps supported on a multiprocessor of the GPU.
It can obtained with CUDA Profiler.} of 97.7\%.

\subsection{Barnes-Hut Algorithm}

The All-Pairs algorithm is simple and straightforward for implementation on a GPU and perfectly fits CUDA.
Yet this algorithm is purely scalable. The corresponding computation time grows like $O(N^2)$,
and its performance is very slow  already for an ensemble of $10^5$ particles.

\begin{figure}[tpb]
\center
\includegraphics[width=12cm]{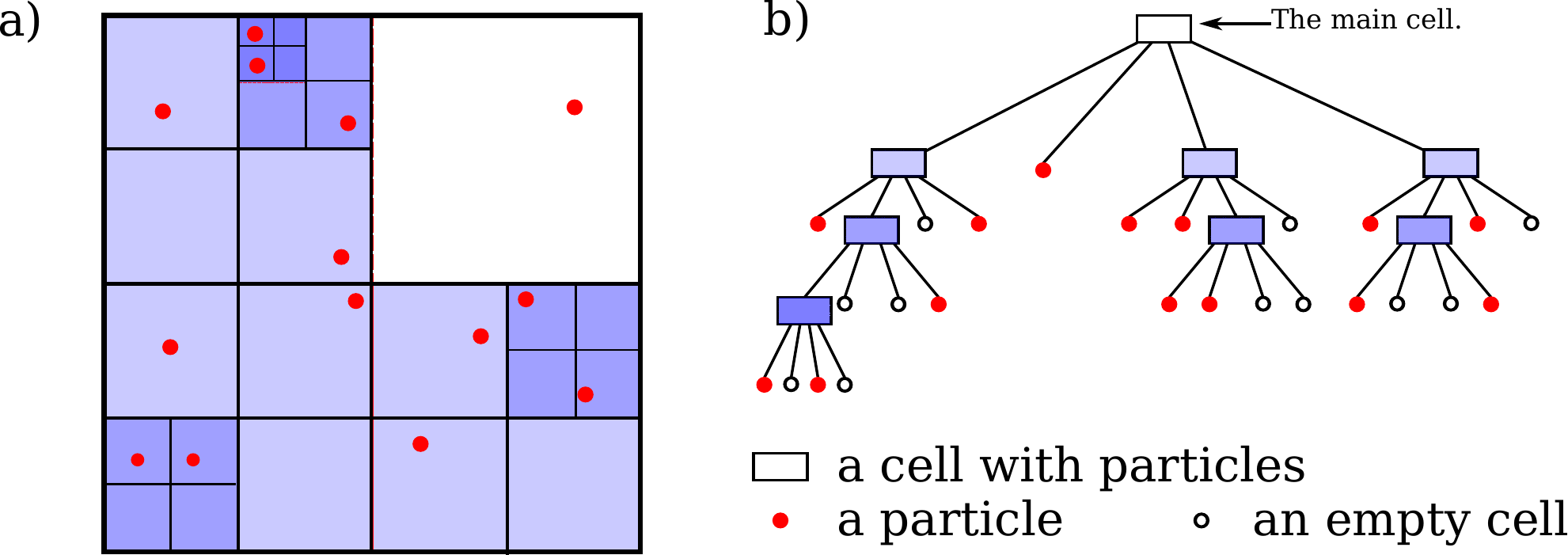}
\caption{Barnes-Hut hierarchical decomposition in two-dimensional space
and the corresponding octree. See Section 3.2 for more details.}
\label{fig:bh}
\end{figure}

The Barnes-Hut (BH) approximation \cite{Barnes1986} exhibits a much better scalability and its computational time
grows like $O(N \log N)$. The key idea of the algorithm is to substitute
a group of particles with a single pseudo-particle mimicking the action of the group.
Then the force exerted by the group on the considered particle can be replaced with the force exerted by the pseudo-particle.
In order to illustrate the idea assume that all particles are located in a three-dimensional cube,
which is named 'main cell'. The main cell is divided then into eight sub-cells. Each
sub-cell confines subset of particles. If there are more than one particle in
the given sub-cell then the last is again divided into eight sub-cells. The procedure is re-iterated until there
is only one particle or none left in each sub-cell. In this way we can obtain an octree with
leaves that are either empty or contain single particle only. A simplified, two-dimensional realization of this algorithm is sketched with
Fig. \ref{fig:bh}. By following this recipe, we can assign to every cell, obtained during the decomposition,
a pseudo-particle, with magnetic moment $\vec
{u'}$ equals to the sum of magnetization vectors $\vec u$ of all
particles belonging to the cell, and position $\vec {\rho'}$ which is the position of the set geometric center,
\begin{equation}
 \label{eq:magCenter}
\vec {\rho'} =
\frac {\sum\limits_{i = 1}^{N'} {\vec {\rho}_i}} {N'},
\end{equation}
where $N'$ is a number of particles in the cell and $\vec {\rho}_i$ is the position of the
$i$-th particle from the sub-set.

Forces acting on $k$-th particle can be calculated by
traversing the octree. If the distance from the particle to the pseudo-particle
that corresponds to the root cell is large enough, the influence of this
pseudo-particle on the $k$-th particle is calculated; otherwise pseudo-particles of
the next sub-cells are checked etc (sometime this procedure can lead finally to a leaf with only one particle in the
cell left). Thus calculated force is added then to the total force acting on the $k$-th particle.

The BH algorithm allows for a high parallelism and it is widely employed in
computational astrophysics problems \cite{Barnes1994}.
However, implementation of the Barnes-Hut algorithm on GPUs remained a challenge until recently, because
the procedure uses an irregular tree structure that does not fit the CUDA architecture well. It is the main reason
why the BH scheme was not realized entirely on
a GPU but some part of calculations was always delegated and performed on a CPU \cite{Gaburov2010,
Jiang2010}. A realization of the algorithm solely on a GPU has been proposed in 2011 \cite{Burtscher2011}.
Below we briefly outline the main idea and specify its difference from the standard, CPU-based realization.


To build an octree on a CPU usually heap objects are used. 
These objects contain both child-pointer and data fields, and
their children are dynamically allocated. To avoid time-consuming dynamic
allocation and accesses to heap objects, an array-based data structure should be used. Since we have several arrays
responsible for variables, the coalesced global memory access is possible.
Particles and cells can have the same data fields, e.g. positions. In this case the same arrays are used.

In contrast to the All-Pairs algorithm, where only two kernels were involved, in the original GPU-BH algorithm has six kernels \cite{Burtscher2011}:
\begin{enumerate}
 \item {Bounding box definition kernel.}
 \item {Octree building kernel.}
 \item {Computing geometrical center and total magnetic moment of each cell.}
 \item {Sorting of the particles with respect to their positions.}
 \item {Computing forces and fields acting on each particle.}
 \item {Integration kernel.}
\end{enumerate}

Kernel 1 defines the boundaries of the root cell. Though 
the ensemble is confined to a container and particles
cannot go outside, we keep this kernel. The size of the root cells can be significantly
smaller than the characteristic size of the container. Moreover, the computation time of this kernel is
very small, typically much less than 1\% of the total time of one integration step.
The idea of this kernel is to find minimum and maximum values of particle positions.
Here we use atomic operations and built-in $min$ and $max$ functions.

Kernel 2 performs hierarchical decomposition of the root cell and builds an octree in the three-dimensional case. 
As well as in following kernels, the particles are assigned to the threads in round-robin fashion.
When a particle is assigned to a thread, it tries to lock the appropriate child pointer. In the case of success, the thread rewrites
the child pointer and releases a lock. To perform a lightweight lock, that is used to avoid several threads access to the same 
part of the tree array, atomic operation should be involved. 
To synchronize the tree-building process we use the ${\_\_syncthreads}$ barrier.

Kernel 3 calculates magnetic moments and positions of pseudo-particles associated with cells by
traversing un-balanced octree from the bottom up. A thread checks if magnetic moment and geometric center of 
all the sub-cells assigned to its cell have already been computed. If not then the thread updates 
the contribution of the ready cells and waits for the rest of the sub-cells. Otherwise the impact of all 
sub-cells is computed.

Kernel 4 sorts particles in accordance to their locations. This step can speed up the performance of 
the next kernel due to the optimal global memory access.

Kernel 5 first calculates forces acting on the particles, and then calculates the corresponding increments. 
Then,in order to compute the force and dipole field
acting upon the particle, the octree is traversed. To minimize thread divergence, it is very important that spatially close
particles belong to the same warp. In this case the threads within the warp would traverse approximately the same
tree branches. This has already been provided by kernel 4. The necessary data
to compute interaction are fetched to the shared memory by the first thread of a warp.
This allows to reduce number of memory accesses.

Finally, kernel 6 updates the state of the particles by using the position increments and re-orient particle magnetic moments.

The above-described algorithm has many advantages. Among them 
are minimal thread divergence and the complete absence of GPU/CPU data transfer (aside of the transfer of the final results), 
optimal use of global memory with minimal number of accesses, data field re-use, minimal number of locks etc.
All this allows to achieve a tangible speed-up. For more detailed 
description we direct the interested reader to Ref. \cite{Burtscher2011}.


\section{Results}
\label{sc:results}
We performed simulations on (i) a PC with Intel Xeon x5670 @2.93GHz CPU(48 Gb RAM) and (ii) a Tesla M2050 GPU.
Though the CPU has six cores, only a
single core was used in simulations. The programs were compiled with \texttt{nvcc}
(version 4.0) and \texttt{gcc} (version 4.4.1) compilers. Since there was no need in high-precision calculations,
we used single-precision variables (\texttt{float}) and compiled the program with
\texttt{-use\_fast\_math} key. We also used \texttt{-O3} optimization flag to speedup our programs.
Finally, the Euler-Maruyama method with time
step $\Delta\tau = 0.001$ was used to integrate Eqs. (\ref{eq:red_mot_rot_tet} -
\ref{eq:red_mot_disp_final}).

We measured the
computation time of one integration step for both algorithms as functions of $N$. The results are presented with Table
\ref{tbl:CompTimeTable}. The benefits of the GPU computing increase with the number of particles. For an ensemble of
$N = 10^6$ particles the speed-up gained from the use of the Barnes-Hut algorithm is almost $300$ compared to the performance
of the performance of the optimized All-Pairs algorithm on the same GPU. However, for $N = 10^3$ the All-Pairs algorithm performs better. It
is because the computational expenses for the tree-building phase, sorting etc., overweight
the speed-up effect of the approximation for small number of particles. Here we remind that $N = 10^3$ was the typical scale of the most of
ferrofluid simulations to date \cite{Wang2002, AA2007}.

\begin{table}
 \centering

 \begin{tabular}{|c|c|c|c|c|c|c|c|}

 \hline
$N$ & $AP_{CPU}$ & $BH_{CPU}$ & $AP_{GPU}$ & $BH_{GPU}$ & $\frac{AP_{CPU}}{AP_{GPU}}$ &
$\frac{AP_{CPU}}{BH_{GPU}}$ & $\frac{AP_{GPU}}{BH_{GPU}}$\\
\hline
$10^3$ & 34 & 9.8 & 0.7 & 2 & 49 & 17 & 0.35\\
$10^4$ & 3 470 & 137 & 20 & 6.5 & 174 & 534 & 3.1\\
$10^5$ & 392 000 & 2 487 &1 830 & 54 & 214 & 7 259 & 33.9\\
$10^6$ & 39 281 250 & 73 121 & 184 330 & 621 & 213 & 63 214 & 297\\
\hline

\end{tabular}
\caption{Duration of single integration step (ms) for the optimized All-Pairs algorithm
implemented on CPU ($AP_{CPU}$) and GPU ($AP_{GPU}$), and for the CPU- ($BH_{CPU}$) 
and GPU-oriented ($BH_{GPU}$) Barnes-Hut algorithm.}
\label{tbl:CompTimeTable}
\end{table}

\begin{figure}[H]
\begin{center}
\includegraphics[scale=0.7]{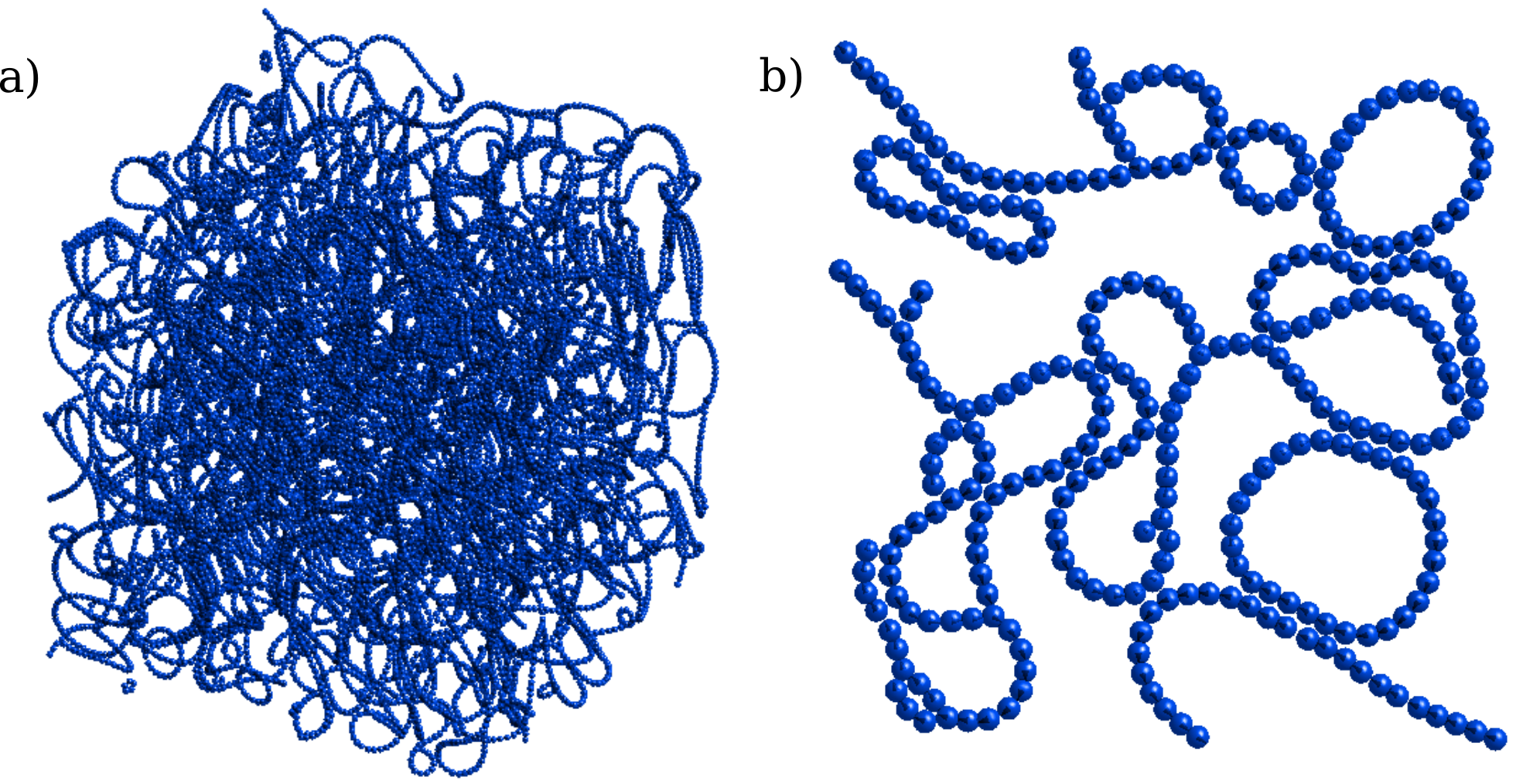}
\end{center}
\caption{Snapshots of $N$-particle ensembles obtained with the
Barnes-Hut algorithm: (a) $N =
20\:000$ (cubic confinement with the edge length $L = 150R$) and (b)
mono-layer of $N = 300$ particles. The parameters are:
$\varGamma_r = \varGamma_d = 0.1$, $T = 300$ K, $\mu = 3.1
\cdot 10^5$ A/m, $D = 5000$ $\rm {kg/m^3}$, $R = 10$ nm.}
\label{fig:FFLSim}
\end{figure}

Fig. \ref{fig:FFLSim} shows instantaneous configurations
obtained during the simulations for a cubic confinement and for a mono-layer.
To simulate a mono-layer of particles we use a rectangular parallelepiped of the height $2.1R$
as a confinement. The parameters of the simulations correspond to the regime when the
average dipole energy is much larger than the energy of thermal fluctuations.
The formation of chain-like large-scale clusters \cite{Shliomis1974} is
clearly visible.

\section{A benchmark test: average magnetization curves}

In order to check the accuracy of the numerical schemes we calculated the reduced magnetization
curve \cite{Shliomis1974}. Reduced magnetization vector is given by the sum  $\langle \vec u \rangle
= 1 / N \sum\limits_{i = 1}^{N} {\vec u_i}$. The main parameters that
characterize ferrofluid magnetic properties are the dipole coupling constant $\lambda$, which is the
ratio of dipole-dipole potential and thermal energy, namely
\cite{Wang2009},
\begin{equation}
\lambda = \frac{\mu_0 m^2}{16 \pi R^3 k_{\rm B} T},
\label{eq:lambda}
\end{equation}
and the volume fraction, which is the ratio between the volume occupied by particles and the total
volume occupied by the ferrofluid, $V_f$, i.e., 
\begin{equation}
\phi = \frac{4/3 \pi R^3 N}{V_f} \cdot 100 \%.
\label{eq:phi}
\end{equation}
In the limit when $\lambda < 1$ and for
small volume fraction, $\phi \ll 100\%$, the projection of the reduced magnetization vector on the direction of applied field,
$\langle u_{\vec H} \rangle$, can well be approximated by the Langevin function
\cite{Shliomis1974}:
\begin{equation}
\langle u_{\vec H} \rangle = L(\alpha) = \coth(\alpha) - \frac{1}{\alpha},
\label{eq:Langevin}
\end{equation}
where $\alpha$ denotes the ratio between magnetic energy and thermal energy, 
\begin{equation}
\alpha =
m \mu_0 H / k_{\rm B} T.
\label{eq:alpha}
\end{equation}

We simulated a system with the parameters corresponding to maghemite
($\gamma-\rm{Fe_2O_3}$) with a saturation magnetization of $\mu = 3.1 \cdot 10^5$ A/m and density $D = 5000$ $\rm {kg/m^3}$,
the carrier viscosity $\eta = 0.89 \cdot 10^{-3}$ Pa (the latter corresponds to the water viscosity at
$T = 298$ K). The volume
fraction is set at $\phi = 1\%$ and the particle radius $R = 3$ nm. The external magnetic
field $\vec H$ was applied along $z-$axis. We initiated the system at
time $\tau = 0$ by randomly distributing particles in a cubic
container. The orientation angles of particle magnetization vectors were obtained by drawing random
values from the interval $[0, 2\pi]$.

\begin{figure}[H]
\begin{center}
\includegraphics[scale=0.7]{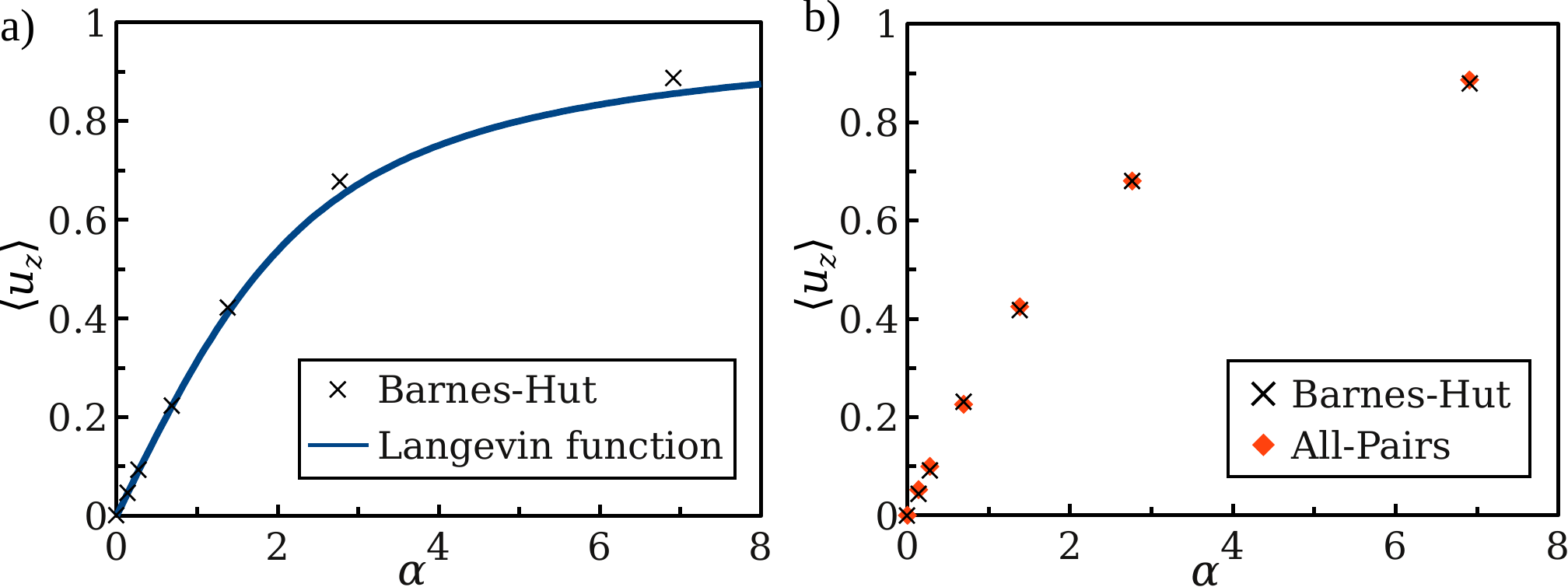}
\end{center}
\caption{Checking the Barnes-Hut approximation: a) comparison between the single-run results obtained with
$N = 10^5$ particles and Langevin function, Eq. (\ref{eq:Langevin}); b) comparison of the results obtained with the
All-Pair and the Barnes-Hut algorithms for an ensemble of $N = 10^3$ particles.}
\label{fig:Langevin}
\end{figure}

\begin{figure}[H]
\begin{center}
\includegraphics[scale=0.7]{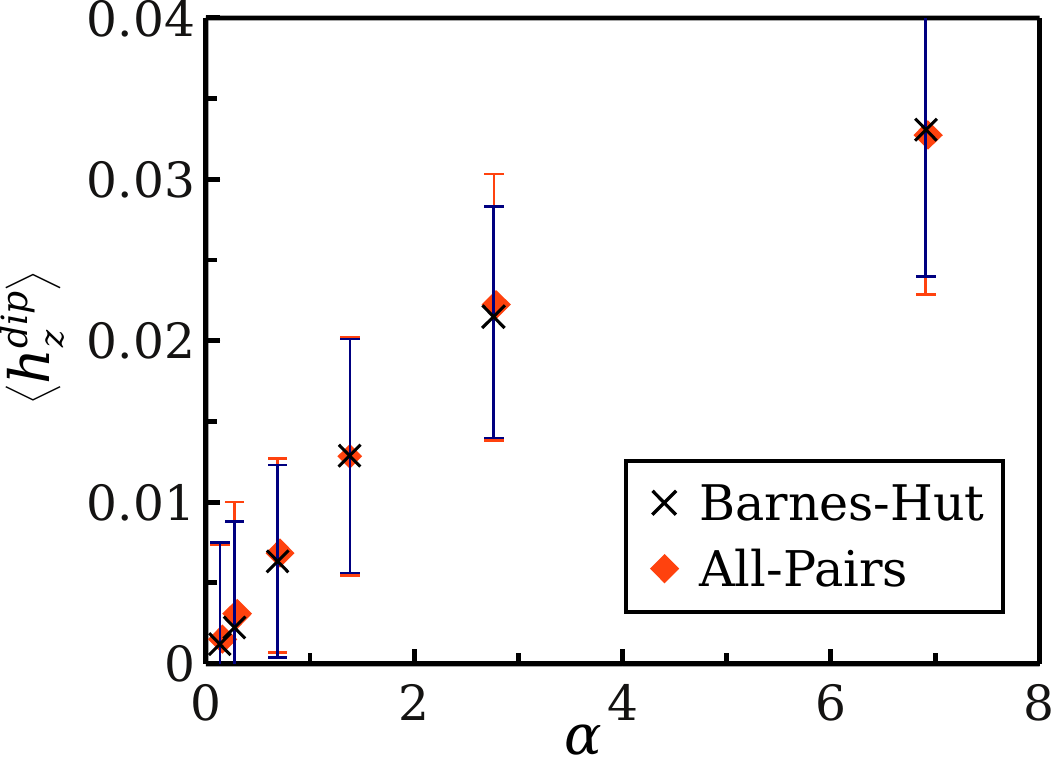}
\end{center}
\caption{$z$-component of the reduced average dipole field as a function of $\alpha$ \cite{AA2007}.
The parameters are the same as in Fig. \ref{fig:Langevin}(b). Each point was obtained by averaging over $10^3$ independent realizations.}
\label{fig:hDip}
\end{figure}

Fig. \ref{fig:Langevin} presents the results of the simulations. After the transient $\tau_{eq} = 1000$, given to the system of
$N = 10^5$ particles to equilibrate, the mean reduced magnetization has been calculated by averaging $\langle u_z\rangle$ over the time
interval $\tau_{calc} = 1000$. It is noteworthy that even single-run results are very close to the Langevin function,
see Fig. \ref{fig:Langevin}(a). The contribution of the magnetostatic
energy grows with $\alpha$ so that the strength of the dipole-dipole interaction is also
increasing, see Fig. \ref{fig:hDip}.

Since the average value of dipole field projection on $z$ axis is positive and increases with $\alpha$, the dipole field amplifies 
the external magnetic field. This explains the discrepancy between the analytical and numerical results obtained for large values of $\alpha$.
For an ensemble of $N = 10^3$ particles the results obtained with two algorithms are near identical,
Fig. \ref{fig:Langevin} (b).

It is important also to compare the average dipole fields, $\langle h^{dip} \rangle$, calculated with the Barnes-Hut and the
All-Pairs algorithms.
The outputs obtained for the above-given set of parameters
are shown on Fig.\ref{fig:hDip}. Again, two algorithms produced almost identical results.

\section{Conclusions}

With this work we demonstrate that the Barnes-Hut algorithm can be efficiently
implemented for large-scale, GPU-based ferrofluid simulations. Overall, we achieved a
speed-up more than two orders of magnitude when compared to
the performance of a common GPU-oriented All-Pairs algorithm. The proposed approach allows to
increase the size of ensembles by two orders of magnitude compared to the present-day scale of simulations \cite{Cerda2010}.
The Barnes-Hut algorithm correctly accounts for the dipole-dipole
interaction within an ensemble of $N = 10^6$ particles and produces results that fit the theoretical predictions with high accuracy.
Our finding opens several interesting perspectives. First, it brings about possibilities to perform
large-scale molecular-dynamics simulations for the time evolution of ferrofluids
placed in confinements of complex shapes like thin vessels or tangled pipes,
where the boundary effects play an important role \cite{Wang2003}.
It is also possible to explore the \textit{non-equilibrium} dynamics of ferrofluids,
for example, their response to different types of externally applied magnetic
fields, such as periodically alternating fields \cite{Mahr1998}, or gradient fields
\cite{erb2008}. Another direction for further studies is the exploration of the relationship between
shape and topology of nano-clusters and different macroscopic
properties of ferrofluids \cite{mendelev2004, borin2011}. Finally, the GPU-based computational algorithms 
provide with new possibilities to study heat transport processes in ferrofluids \cite{Ganguly2004}, for example to investigate 
the performance of ferromagnetic particles as heat sources for magnetic fluid hyperthermia \cite{rosensweig2002}.

\section{Acknowledgment}
A.Yu.P. and T.V.L. acknowledge the support of the Cabinet of Ministers of
Ukraine obtained within the Program of Studying and Training for Students, PhD Students and
Professor's Stuff Abroad and the support
of the Ministry of Education, Science, Youth, and Sport of Ukraine (Project No
0112U001383). S.D. and P.H. acknowledge the support by the cluster of excellence
Nanosystems Initiative Munich (NIM).

\bibliographystyle{model1-num-names}


\end{document}